\documentclass[prb,aps,graphicx,color,float]{revtex4}
\usepackage{epsfig}
\usepackage{bm}

\begin{document}

\preprint{}

\newcommand{\stef}[2]{$\blacktriangleright${\sc round #1:}{\em #2}$\blacktriangleleft$}

\title{Detecting non-Abelian Statistics with Electronic Mach-Zehnder
Interferometer}

\author{D. E. Feldman$^1$ and Alexei Kitaev$^{2,3}$}
\affiliation{$^1$Department of Physics, Brown University, Providence,
Rhode Island 02912, USA\\
$^2$California Institute of Technology, Pasadena, California 91125, USA}
\affiliation{$^3$Microsoft Project Q, University of California, Santa Barbara,
California 93106, USA}

\begin{abstract}

Fractionally charged quasiparticles in the quantum Hall state with filling
factor $\nu=5/2$ are expected to obey non-Abelian statistics. We demonstrate
that their statistics can be probed by transport measurements in an
electronic Mach-Zehnder interferometer. The tunneling current through the
interferometer exhibits a characteristic dependence on the magnetic flux and a
non-analytic dependence on the tunneling amplitudes which can be controlled by
gate voltages.

\end{abstract}


\maketitle


One of the central features of the quantum Hall effect (QHE) is the fractional
charge and statistics of quasiparticles. The quantum state of bosons or
fermions does not change when one particle makes a full turn around
another. On the other hand, Laughlin quasiparticles pick up nontrivial phases
when they encircle each other. Non-Abelian statistics predicted in some QHE
systems~\cite{MooreRead,nw} is even more interesting: the state vector changes
its direction in the Hilbert space after a particle makes a full circle.

Shot noise experiments~\cite{weizmann-G} allowed the observation of fractional
charges in QHE liquids. Probing fractional statistics is more difficult. It
was argued that the mutual statistics of non-identical quasiparticles was
detected in a recent experiment~\cite{goldman2005} by Camino {\it et al.},
however, the interpretation of the experimental results remains controversial
\cite{kim,goldman2006}. There are several theoretical proposals for observing
the statistics of identical Abelian quasiparticles~\cite{cfksw,HBT,kane,lfg}
but neither of them has been realized experimentally.

Detecting non-Abelian anyons is of special interest due to their promise for
fault-tolerant quantum computation~\cite{Kitaev97}. One approach for their
observation and probing their statistics is based on current noise in complex
geometries~\cite{bena-nayak,gss}. A simpler proposal involves current through an
Aharonov-Bohm interferometer with trapped quasiparticles~\cite{BKS,SH}. This
method should work if the number of trapped quasiparticles does not fluctuate
on the measurement time scale~\cite{kane}. Such a condition might be difficult
to satisfy in non-Abelian systems, where the excitation gap is relatively
low~\cite{gap_5/2}.

We suggest another method, which is free from this limitation. It uses the
electronic Mach-Zehnder interferometer recently designed at Weizmann
Institute~\cite{MZ}, see Fig.~1(a).  We consider the $\nu=5/2$ QHE state and
show that the tunneling current $I$ through the interferometer contains
signatures of non-Abelian statistics. The current~(\ref{current}) is a
periodic function of the magnetic flux $\Phi$ through the interferometer with
period $\Phi_0=hc/e$, but, in contrast with the Abelian case~\cite{lfg}, it is
not a simple sine wave. Note the non-analytic dependence on the interedge
tunneling amplitudes $\Gamma_1$, $\Gamma_2$ at the quantum point contacts
(constrictions) QPC1, QPC2. In the limit of $\Gamma_2\ll\Gamma_1$ the formula
for the current assumes a sinusoidal form,
$I=I_0+I_\phi\cos(2\pi\Phi/\Phi_0+{\rm const})$, where the flux-dependent and
flux-independent terms are related by the scaling law:
\begin{equation}
\label{1}
I_\Phi(\Gamma_1,\Gamma_2)\sim [I_0(\Gamma_1,\Gamma_2)-I_0(\Gamma_1,0)]^b
\end{equation}
with $b=2$. This can be compared with the case of Fermi statistics, where
$b=1/2$; for Abelian anyons $b=m+1/2$ with an integer $m>0$,
Ref.~\onlinecite{lfg}.

The Letter is organized as follows. First, we briefly discuss the relevant
properties of the $\nu=5/2$ QHE state and the structure of the Mach-Zehnder
interferometer. Next, we derive the above results using a kinetic
equation. Finally, we show how to obtain such an equation for an arbitrary
non-Abelian state.

Elementary excitations in the $\nu=5/2$ liquid carry charge $\pm e/4$. Due to
the non-Abelian statistics, the state of a system of several quasiparticles is
not uniquely determined by the quasiparticle coordinates. It is convenient to
classify the states according to their superselection sectors. (For a review
of this formalism, see Appendix~E in Ref.~\onlinecite{kitaev}.) A
superselection sector is characterized by the electric charge, $q=ne/4$ as
well as ``topological charge'' taking on three values~\cite{SH}: $1$
(``vacuum''), $\epsilon$ (``fermion''), and $\sigma$ (``vortex''). If $n$ is
even then the topological charge can be either $1$ or $\epsilon$; if $n$ is
odd then the topological charge is $\sigma$. The topological charge obeys
these fusion rules which allow the calculation of the topological charge of
the composite system from the charges of its parts:
$\epsilon\times\epsilon=1$,\, $\epsilon\times\sigma=\sigma$,\,
$\sigma\times\sigma=1+\epsilon$.

When a $+e/4$ quasiparticle encircles a composite excitation in the sector
$a=(ne/4,\alpha)$ (where $\alpha$ is the topological charge), it picks up some
statistical phase, which depends not only on $n$ and $\alpha$ but also on the
topological charge $\beta$ of the whole system. We denote this phase by
$\phi_{ab}$, where $b=((n+1)e/4,\beta)$. It is given by the formula
\begin{equation}
\label{2}
\phi_{ab}=n\pi/4+\phi'_{\alpha\beta}.
\end{equation}
where the non-Abelian part $\phi'_{\alpha\beta}$ equals $0$ if the excitation
is in the vacuum sector, $\pi$ if it is in the $\epsilon$ sector, $-\pi/4$ if
the whole system in the vacuum sector, and $3\pi/4$ if the whole system is in
the $\epsilon$ sector. (In all four cases, the other topological charge is
$\sigma$.) We will see that these phases determine the current through the
Mach-Zehnder interferometer.

The interferometer~\cite{MZ} is sketched in Fig.~1(a). Charge propagates along
two chiral edges in the direction shown by arrows. Quasiparticles tunnel
between the edges at the point contacts QPC1 and QPC2. We are interested in
the tunneling current between the edges (the current from source S1 to drain
D2). It depends on the voltage $V$ between the edges (i.e., the difference of
the electrochemical potentials between sources S1 and S2) and the magnetic
flux $\Phi$ through the loop A-QPC2-B-QPC1-A. This loop is defined so as not
to touch the leads, because we assume that the leads fully absorb edge
excitations. We also assume that the tunneling amplitudes $\Gamma_1$ and
$\Gamma_2$ are small, therefore we can use perturbation theory. In physical
terms, individual tunneling events are regarded as independent and assigned
certain probabilities, which are calculated below. When a $+e/4$ quasiparticle
tunnels from the outer edge to the inner edge (through QPC1 or QPC2), the
electric charge on the inner edge increases by $e/4$ (cf.\
Ref.~\onlinecite{lfg}), and the topological charge changes according to the
fusion rules~\cite{footnote1}.  Specifically, the initial value of the
topological charge fuses with $\sigma$: if the initial charge is $1$ or
$\epsilon$ then the final charge is $\sigma$; if the initial charge is
$\sigma$ then the final charge is $1$ or $\epsilon$. In the latter case, the
two fusion outcomes occur with equal probabilities: $P^{+}_{\sigma\to
1}=P^{+}_{\sigma\to\epsilon}=1/2$. Indeed, the tunneling process is
independent of the global edge state (due to the absorbing properties of the
leads), hence we may assume that the fusing charges come from uncorrelated
sources: the inner edge forms a topologically neutral object with the outer
edge, and the tunneling quasiparticle is part of a particle-antiparticle pair
created from the vacuum. The probabilities are calculated by applying a
topological charge operator of the subsystem {\bf (inner edge + tunneling
quasiparticle)} to the four-body state described above as discussed in
Ref. \onlinecite{kitaev} and footnote \onlinecite{footnote2}.  
A similar argument applies to $-e/4$
quasiparticles or, equivalently, to the tunneling of $+e/4$ quasiparticles
from the inner to the outer edge. We will use the notation $P^{-}_{a\to b}$ in
this case, though the superscript turns out to be redundant. To summarize,
$P^{\pm}_{1\to\sigma}=P^{\pm}_{\epsilon\to\sigma}=1$,\, $P^{\pm}_{\sigma\to
1}=P^{\pm}_{\sigma\to\epsilon}=1/2$, all other probabilities being zero.

Since bulk excitations are gapped, the low-energy physics is determined by
edges~\cite{wen}. Thus, the Hamiltonian is
\begin{equation}
\label{3}
\hat H=\hat H_{\rm edge}
+[e^{-i\omega t}(\Gamma_{1}\hat X_1 + \Gamma_{2}\hat X_2) + H.c.],
\end{equation}
where $\hat H_{\rm edge}$ is the Hamiltonians of the two edges (which carry
opposite topological charges but are otherwise independent),
$\omega=eV/4\hbar$ describes the voltage bias (cf. Ref. \onlinecite{lfg}), and
the operators $\hat X_1$, $\hat X_2$ correspond to the $e/4$ charge transfer
from the outer to the inner edge at QPC1 and QPC2, respectively. The forward
and backward tunneling rates, $w^{+}$ and $w^{-}$ can be calculated in the
second order of perturbation theory.  It is convenient to consider first the
hypothetical setup shown in Fig.~1(b), where the current is independent of the
topological charges and the magnetic flux.  This yields an expression of the
form $\sum_{j,k}r^{\pm}_{jk}\Gamma_j^*\Gamma_k$.  In the real problem, the
rates also depend on the superselection sector of the inner edge in the
initial and final state~\cite{footnote3}. For example,
\begin{equation}
r^{+}_{12}(a\to b)\,=\,\hbar^{-2}\int_{-\infty}^{+\infty}
\langle X_{1}^{\dag}(t)\Pi_{b}X_{2}(0)\rangle_{a}\,e^{i\omega t}\,dt,
\end{equation}
where $\langle\cdots\rangle_a$ denotes the thermodynamic average restricted to
the superselection sector $a$, and $\Pi_b$ is the projector onto the sector
$b$. The integrand has the meaning of a particle tunneling through QPC2 and
returning via QPC1, hence it incorporates both the Aharonov-Bohm phase due to
the magnetic field, $\phi_{\rm mag}=2\pi\Phi/(4\Phi_{0})$ and the statistical
phase $\phi_{ab}$, Eq. (\ref{2}). 
It also includes the fusion probability $P^{+}_{a\to
b}$. Summing over the four possible paths, we obtain this result:
$w^{\pm}_{a\to b}=P^{\pm}_{a\to b}u^{\pm}_{a\to b}$, where
\begin{equation}
\label{5}
u^{+}_{a\rightarrow b}=r_{11}^{+}\left(|\Gamma_1|^2+|\Gamma_2|^2\right)
+\left(r_{12}^{+}e^{i\phi_{\rm mag}}e^{i\phi_{ab}}\Gamma_1^*\Gamma_2
+c.c.\right).
\end{equation}
The back tunneling rate can be obtained from the detailed balance principle,
$u^{-}_{b\to a}=\exp[-eV/(4k_B T)]\,u^{+}_{a\to b}$. The fusion probabilities
$P^{+}_{a\to b}$ also satisfy a detailed balance equation, see
Eq.~(\ref{dbalance}). The constants $r_{11}^{+}$ and $r_{12}^{+}$ in~(\ref{5})
depend on the expressions for the operators $\hat H_{\rm edge}$, $\hat X_1$
and $\hat X_2$, as well as on the temperature $T$ and the voltage $V$. They
are independent of the magnetic flux through the interferometer and the charge
labels $a$, $b$ and could be calculated using the simplified geometry of
Fig.~1(b). A calculation based on Wen's hydrodynamic model~\cite{wen-model}
will be published elsewhere. In this Letter we discuss those properties which
are not sensitive to edge physics details, but rather to the fractional
statistics of tunneling particles.

The statistical phase factor $\exp(i\phi_{ab})$, Eq.~(\ref{2}), is invariant
under the fusion of both $a=(ne/4,\alpha)$ and $b=((n+1)e/4,\beta)$ with an
electron, whose superselection sector is $(e,\epsilon)$. Thus, the
superselection sectors form 6 equivalence classes, which are characterized by
$(n\bmod 4)$ and the choice of $1$ or $\epsilon$ if $n$ is even.  These
classes and possible transitions between them are shown in Fig.~2. At zero
temperature (and positive $V$) the transitions occur only in the direction of
arrows since quasiparticles must tunnel from the edge with higher potential to
the edge with lower potential. At finite temperatures transitions between the
states connected by the lines in Fig.~2 occur in both directions.

The average tunneling current is given by the equation
\begin{equation}
\label{6}
I=\frac{e}{4}\sum_{ab} f_a(w^{+}_{a\rightarrow b}-w^{-}_{a\rightarrow b}),
\end{equation}
where the distribution function $f$ can be found from the steady state
condition
\begin{equation}
\label{7}
\frac{df_a}{dt}=\sum_{b}
\left [ f_b(w^{+}_{b\rightarrow a}+w^{-}_{b\rightarrow a})
- f_a(w^{+}_{a\rightarrow b}+w^{-}_{a\rightarrow b})\right ]=0.
\end{equation}
The solution of the system of linear equations~(\ref{7}) is tedious but
straightforward.  The general expression for the current is lengthy and will
be published elsewhere.  It simplifies at zero temperature:
\begin{equation}
\label{current}
I=\frac{e}{4} r^{+}_{11} \left[|\Gamma_1|^2+|\Gamma_2|^2\right ] 
\frac{1-\lambda^2+\frac{\lambda^4}{8}[1-\cos(2\pi\Phi/\Phi_0+\delta)]}
{1-\frac{3}{4}\lambda^2+\frac{\lambda^4}{16}[1-\cos(2\pi\Phi/\Phi_0+\delta)
-\sin(2\pi\Phi/\Phi_0+\delta)]},
\end{equation}
where $\lambda=|r^{+}_{12}/r^{+}_{11}|\cdot
2|\Gamma_1\Gamma_2|/(|\Gamma_1|^2+|\Gamma_2|^2)$ and $\delta=4\mathop{\rm
arg}(r^{+}_{12}\Gamma_1^*\Gamma_2)$.  The current is the ratio of two linear
trigonometric polynomials of $2\pi\Phi/\Phi_0$. In Abelian quantum Hall states
the current is a sine wave plus a constant~\cite{lfg}. Thus, if the
experimentally measured dependence of the tunneling current on the magnetic
field at fixed voltage can be fitted by an equation of the form
(\ref{current}), this proves non-Abelian statistics of elementary excitations.

The current is a periodic function of the magnetic flux with period
$\Phi_0$. This agrees with the Byers-Yang theorem~\cite{BY}, which applies to
any interferometer with the magnetic flux passing through a hole.

The dependence of the current on the tunneling amplitudes is non-analytic.  If
$\Gamma_2\ll\Gamma_1$ then the current~(\ref{current}) can be expanded in
powers of $\Gamma_2$. This gives Eq.~(\ref{1}), which can be used for another
experimental test of non-Abelian statistics.  The tunneling amplitudes are
controlled by gate voltages.  At fixed values of the gate voltages, one can
measure the magnetic field dependence of the current and extract the
flux-dependent and flux-independent contributions as $I_0=[{\rm max}_\Phi
I(\Phi)+{\rm min}_\Phi I(\Phi)]/2$ and $I_\Phi=[{\rm max}_\Phi I(\Phi)-{\rm
min}_\Phi I(\Phi)]/2$.  Changing the gate voltages will allow the testing of
the scaling relation~(\ref{1}). Eq.~(\ref{1}), however, would have the same
form in the (rather unlikely) situation where a quasiparticle picks up an
Abelian phase of $\pi/2$ after a full circle around another quasiparticle. On
the other hand, the flux dependence (\ref{current}) emerges only in the
non-Abelian case. In contrast to the Abelian case, the I-V curve is asymmetric 
for $\nu=5/2$: at $V<0$ one has to change the overall sign of the current (\ref{current}) {\it and} 
the sign before $\sin(2\pi\Phi/\Phi_0+\delta)$ in the denominator.

So far we have ignored the possibility of quasiparticle trapping inside the
interferometer.  Quasiparticles can tunnel not only between the edges but also
to or from localized states in the electron liquid. In the usual
interferometer geometry~\cite{BKS,SH}, such tunneling events suppress the
interference picture. Indeed, the current through the interferometer depends
on the topological charge between the contacts. Each tunneling event between
an edge and a localized state changes the topological charge. After averaging
with respect to the fluctuating charge the interference picture disappears. On
the other hand, tunneling to localized states plays little role in the
Mach-Zehnder interferometer as long as the typical interval between such
events exceeds the time between two consecutive tunneling events at QPC1 and
QPC2. Indeed, any localized charges may be attributed to either the inner edge
or the outer edge depending on their position relative to the interference
loop. Hence, one can still use Eqs.~(\ref{6},\ref{7}) for the calculation of
the average current, provided those charges are fixed.

Algebraic theory of anyons~\cite{kitaev} allows the calculation of the
tunneling rates $w^{\pm}_{a\rightarrow b}$ for an arbitrary type of
non-Abelian statistics. As discussed above, the calculation of those rates
reduces to the calculation of the statistical phase factors $\exp(i\phi_{ab})$
and the fusion probabilities $P^{\pm}_{a\to b}$. These are given by the
following formulas:

\begin{equation}
\label{9}
\exp(i\phi_{ab})=\frac{\theta_{b}}{\theta_{a}\theta_{x}},\quad
P_{a\to b}^{+}=N^{b}_{ax}\frac{d_{b}}{d_{a}d_{x}},
\end{equation}
where $x$ refers to the tunneling quasiparticle, $\theta_{x}$ and $d_{x}$ are
the topological spin and the quantum dimension, respectively, and $N^{b}_{ax}$
is the fusion multiplicity. The fusion probability $P_{a\to b}^{-}$ is
obtained by replacing $x$ with its antiparticle $\bar{x}$. Note the detailed
balance equation:
\begin{equation}
\label{dbalance}
d_{b}^{2}P^{-}_{b\to a}=d_{a}^{2}P^{+}_{a\to b}.
\end{equation}
In the $\nu=5/2$ Pfaffian state, all fusion multiplicities are equal to $1$,
and the quantum dimensions and topological spins can be found in Table~I
(cf. Ref. \onlinecite{kitaev}). Using that table one can reproduce the above
results for the tunneling probabilities.

In conclusion, we have calculated the tunneling current through the
Mach-Zehnder interferometer for the $\nu=5/2$ QHE liquid. The dependence of
the current on the magnetic flux and tunneling amplitudes can be used for
probing non-Abelian statistics. We thank K.\,T.\,Law for the help with
figures. A.\,K. acknowledges the support by ARO under Grants No.\
W911NF-04-1-0236 and W911NF-05-1-0294, and by NSF under Grant No.\
PHY-0456720. D.\,E.\,F. acknowledges the support by NSF under Grant No.
DMR-0544116 and the hospitality of Microsoft Station Q.

\begin{table}
\vspace{3mm}
\begin{tabular}{|c|c|c|c|}
	\hline
Topological charge &  Electric charge    &   $d_a$  & $\theta_a$     \\
	\hline
          1        & $me/2$              & 1        & $\exp(i\pi m^2/2)$    \\
$\epsilon$ & $me/2$    & 1          & $-\exp(i\pi m^2/2)$\\
$\sigma$ & $e/4+me/2$ & $\sqrt{2}$ &  $\exp(i\pi[2m^2+2m+1]/4)$ \\

	\hline
\end{tabular}
\caption{Statistics in the Pfaffian state.}
\label{table_I}
\vspace{3mm}
\end{table}


\begin{figure}
\centerline{\begin{tabular}{c@{\qquad}c}
\epsfig{file=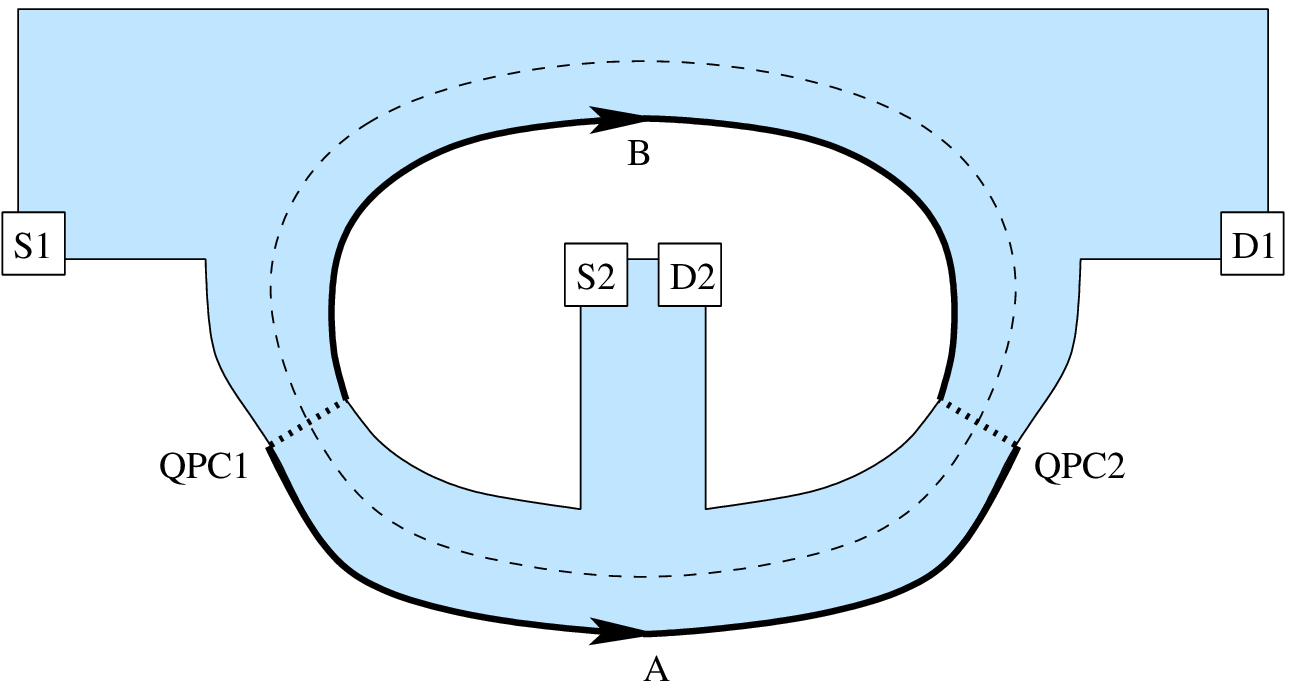, scale=0.75} &
\raisebox{0.25in}{\epsfig{file=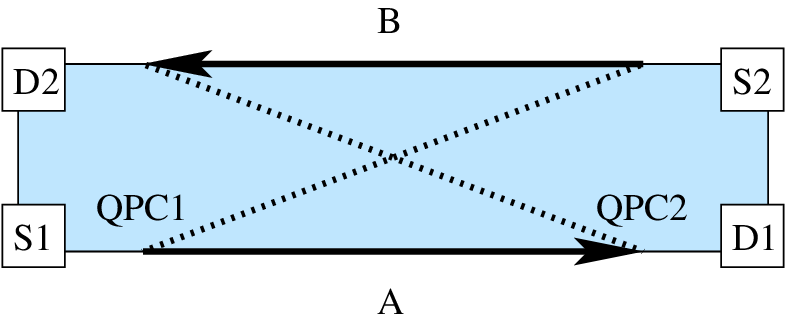, scale=0.75}} \\
a) & b)
\end{tabular}}
\caption{a)~Schematic picture of the Mach-Zehnder interferometer: S1, S2, D1,
D2 denote sources and drains; arrows show the edge mode propagation direction;
quasiparticles tunnel between the edges at quantum point contacts QPC1 and
QPC2. b)~Hypothetical setup for the calculation of normalized tunneling
rates.}
\label{fig1}
\end{figure} 

\begin{figure}
\epsfig{file=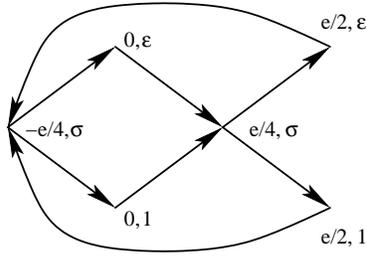, scale=0.75} 

\caption{Six states of the interferometer labeled by the electric and
topological charge on the inner edge. Lines show possible transitions. At
zero temperature the transitions occur only in the direction of arrows.}
\label{fig2}
\end{figure}


\begin{thebibliography} {150}
\bibitem{footnote1} The electric charge on each edge is conserved modulo $e$
since electrons can come and go through the leads. The topological charge is
not affected by the leads and can be, in principle, measured by transporting a
quasiparticle along the dashed line in Fig.~1(a). The charges on the outer edge
are always opposite to those on the inner edge (for the topological charge in the Pfaffian
state \cite{MooreRead} considered in this Letter,
``opposite'' is the same as ``equal'').
\bibitem{footnote2} The relevant degrees of freedom can be characterized by
Majorana operators $c_{1}$, $c_{2}$, $c_{+}$, $c_{-}$ pertaining to edge~1
(outer), edge~2 (inner), the tunneling particle, and its antiparticle. The
initial state $|\Psi\rangle$ is an eigenstate of $ic_{1}c_{2}$ and
$ic_{+}c_{-}$ whereas the topological charge in question is described by the
operator $ic_{2}c_{+}$ (charge $1$ corresponds to eigenvalue $1$, and
$\epsilon$ corresponds to $-1$). It is easy to see that
$\langle\Psi|ic_{2}c_{+}|\Psi\rangle=0$, hence $P_{\sigma\to
1}=P_{\sigma\to\epsilon}$.
\bibitem{footnote3} We neglect quantum interference between states with
different charges on the edges because superpositions of such states decohere
faster than the tunneling rate. This is true as long as the second order
perturbation theory is valid, which can be demonstrated by deriving the
kinetic equation~(\ref{7}) using the Keldysh technique (cf.\
Ref.~\onlinecite{lfg}).
\end{thebibliography}
\end{document}